%% file: lmm_final.tex
\documentclass[doc]{apa}
\usepackage{graphicx,epsfig,amsmath,alltt,setspace,bm,float, apacite, mathtools}
\usepackage[english]{babel}
\usepackage{tikz}
\usepackage[utf8]{inputenc}
\usepackage[authormarkup=none]{changes}
\usetikzlibrary{arrows}
\usepackage{xcolor}
\setremarkmarkup{(#2)}
\usepackage{lipsum}
\definechangesauthor[color=orange]{per}

\title{Score-based Tests for Detecting Heterogeneity in Linear Mixed Models}
\author{\qquad}
\affiliation{\qquad}
\fourauthors{Ting Wang}{Edgar C.\ Merkle}{Joaquin A.\ Anguera}
           {Brandon M.\ Turner}
\fouraffiliations{The American Board of Anesthesiology}
                {University of Missouri}
                {University of California, San Francisco}
                {The Ohio State University}

\abstract{
 Cross-level interactions among fixed effects in 
 linear mixed models (also known as multilevel models) can be 
 complicated by heterogeneity stemming from random 
 effects and residuals. When heterogeneity is present, 
 tests of fixed effects (including cross-level interaction terms) are 
 subject to inflated Type I or Type II error. 
 While the impact of variance change/heterogeneity has been 
 noticed in the literature, few methods have been proposed to detect 
 this heterogeneity in a simple, systematic way. In addition, when 
 heterogeneity among clusters is detected, researchers often wish 
 to know which clusters' variances differed from the others. In this 
 study, we utilize a recently-proposed family of score-based tests to 
 distinguish between cross-level interactions and heterogeneity in 
 variance components, also providing information about specific clusters 
 that exhibit heterogeneity. 
 These score-based tests only require estimation of the null 
 model (when variance homogeneity is assumed to hold), and they have 
 been previously applied to psychometric models to detect measurement 
 invariance.  In this paper, we extend the tests to linear mixed models, 
 allowing us to use the tests to differentiate between interaction 
 and heterogeneity. We detail the tests' 
 implementation and performance via simulation, provide an empirical 
 example of the tests' use in practice, and provide code illustrating 
 the tests' general application.
}

\acknowledgements{This work was supported by National Science
  Foundation grants SES-1061334 and 1460719.  
  Correspondence to Ting Wang.  Email: twb8d@mail.missouri.edu}

\shorttitle{Detecting heterogeneity in mixed models}
\rightheader{Detecting heterogeneity in mixed models}

\let\proglang=\textsf
\let\pkg=\emph
\let\code=\texttt

\mathtoolsset{showonlyrefs=false,showmanualtags=false}

\spacing{1}

\usepackage{Sweave}
\begin{document}
\input{lmm_final-concordance}
\maketitle
\include{Sweave}
\setkeys{Gin}{width=\textwidth}

\section{Introduction}
Cross-level interactions are often of interest for researchers 
adopting linear mixed models (also known as multilevel models or 
hierarchical linear models), which 
specifically refers to the interaction between 
a lower-level (level 1) covariate and a upper-level (level 2) covariate.  
For example, in a cross-sectional study of students within schools, 
we might observe a cross-level interaction between students' 
cognitive ability (level 1) and school climate (level 2).  
In a longitudinal study of repeated observations within participants, 
we might note a cross-level interaction between time (level 1) and 
participant socioeconomic status (level 2).  However, the estimation 
and testing of cross-level interactions in linear mixed models (LMMs) is 
often complicated by the multiple variance terms in the model.  
For example, when a cross-level interaction exists, it may 
not be detected due to heterogeneity in a random effect variance or 
in a residual variance \cite{hal04}. If this heterogeneity is not 
accounted for, it can lead to biased standard error 
estimates \cite{lec14, ver96} and incorrect significance tests for 
fixed parameters \cite{kwo07}.   

Heterogeneous (co-)variances often 
occur in longitudinal studies, 
where the heterogeneity in variance/covariance is observed across individuals. 
For example, recent ecological momentary assessment (EMA)
studies have shown that substantial heterogeneity exists across
individuals in the variance and covariance of emotional
states over time \cite{roc09, tho12, kni13, ebn15}. 
A similar phenomenon occurs in education research, where it is of 
interest to model student achievement over time \cite{loc07}. 
This heterogeneity is often due to unobserved covariates or to confounding 
with other covariates \cite<such as time;>{kov12, bre14}; issues that are 
inevitable in many applied scenarios \cite{loc07}.  Hence, it is important 
to develop accessible methods to detect heterogeneity in the multiple 
variance parameters observed in mixed models.  

Several previous methods test for heterogeneity by building it into 
the model and using a likelihood ratio (LR) test or Kolmogorov-Smirnov 
test \cite{ver96, mcl88, ste74}.
The LR test can be used when heterogeneity can be explained by  
observed variables.  For example, 
\pkg{semtree} \cite{brand13}, 
\pkg{longRpart2} \cite{steg18} and \citeA{abd02} all 
utilize a series of LR test to 
detect potential split point(s) along auxiliary variables.  
However, this test can be cumbersome when the variable has many categories, 
and it can be sub-optimal when the variable is ordinal or continuous.
In contrast, the Kolmogorov-Smirnov test utilizes a mixture 
model framework, and checks 
the correct number of mixture components in an omnibus goodness-of-fit test. 
However, as shown by \citeA{ver96}, this approach is subject to low power, 
even decreasing to zero when the residual variance is large.  
 
In this paper, we aim to extend a family of score-based 
tests \cite<e.g.,>{zeihor07,merzei13} to the study of heterogeneity in linear mixed models.  
These tests have been previously applied to 
detect measurement invariance in factor 
analysis \cite{merzei13, MerFanZei, WanMerZei14} and in models from item response 
theory \cite<IRT; e.g., >{strobl14, wang18}.  
This study serves as a novel application of the tests in the 
context of LMM by providing a unified new approach to differentiate 
heterogeneity in variance components (either random effects or residual)
from cross-level interactions.  From a technical perspective, 
this study differs from the previous measurement invariance studies in 
that the casewise scores are not \emph{i.i.d.} any more by definition, 
since the observations within the same cluster are correlated. 
In addition, we focus on heterogeneity in both fixed parameters and
variance/covariance parameters, whereas the related approach of 
\citeA{fok18} has tested only for changes in fixed parameters while 
maintaining homogeneity in variance and covariance parameters.  
We also discuss graphical methods associated with the tests, 
which can be helpful for identifying clusters that exhibit 
similar variance estimates.  In the following section, 
we provide a brief overview of the score-based tests' generalizations to 
linear mixed models.  Next, we report on the 
results of a simulation to examine the 
tests' abilities in the context of linear mixed models.  
Finally, we provide an empirical example with illustrating code
and discuss the tests' future generalizations.  

\section{Linear Mixed Model}
The linear mixed model (LMM) can be expressed in both conditional and 
marginal forms.  The former facilitates theoretical understanding, and 
the latter simplifies the computational expression.  
We will detail these two expressions in 
the following sections, focusing on a two-level model where individual 
observations are nested within a series of clusters.

\subsection{Conditional Expression}
The conditional version of the LMM can be written as
\begin{eqnarray}
  \label{eq:lmmcond}
    \bm y_j |\bm b_j &\sim& N(\bm X_j \bm \beta+\bm Z_j\bm b_j, \bm R_j)\\
  \label{eq:lmmran}
    \bm b_j &\sim& N(\bm 0, \bm D)\\
  \label{eq:lmmres}
    \bm R_j &=& \sigma_{r}^2\bm I_{n_j},
\end{eqnarray}
where $\bm y_j$ is the observed data vector for the $j$th cluster, 
$j=1,\ldots,J$ (so that the level 1 sample size is given 
as $n = \sum_{j=1}^{J} n_j$); $\bm X_j$ is an
$n_j \times p$ matrix of fixed covariates; $\bm \beta$ is the fixed effect
vector of length $p$; $\bm Z_j$ is an $n_j \times q$ design matrix of random
effects; and $\bm b_j$ is the random effect vector of
length $q$.

The vector $\bm b_j$ is assumed to follow a normal distribution with
mean $\bm 0$ and covariance matrix $\bm D$, where $\bm D$ is a 
matrix composed of variances/covariances for random effect parameters.
The residual covariance matrix, $\bm{R}_j$, is the product of the residual
variance $\sigma_{r}^2$ and an identity matrix of dimension $n_j$. Later, the 
matrix $\bm R$ will include residuals across all clusters, so the identity 
matrix is of dimension $n$.

Based on the notations above, the following notation is used to represent 
data and parameters across all clusters in the data.
\begin{eqnarray}
  \label{eq:redefine1}
  \bm y &=& \{\bm y_1, \bm y_2, \ldots, \bm y_j, \ldots, \bm y_J \}\\
  \label{eq:redefine2}
  \bm X &=& \{\bm X_1, \bm X_2, \ldots, \bm X_j, \ldots, \bm X_J \}\\
  \label{eq:redefine3}
  \bm Z &=& \{\bm Z_1, \bm Z_2, \ldots, \bm Z_j, \ldots, \bm Z_J \}\\
  \label{eq:redefine4}
  \bm b &=& \{\bm b_1, \bm b_2, \ldots, \bm b_j, \ldots, \bm b_J \}\\
  \bm G &=& \bm D \otimes \bm I_{J},
\end{eqnarray}
where $\otimes$ is the Kronecker product.

Finally, we define $\bm \sigma^2$ to be a vector of length $K$, 
containing all variance/covariance parameters (including those of the random
effects and the residual). This implies that the matrix $\bm D$ 
has $(K-1)$ unique elements.  For example, in a model with two random effects 
that are allowed to covary, $\bm \sigma^{2}$ is a vector of 
length 4 (i.e., $K = 4$).  The first three elements correspond to the unique
entries of $\bm D$, which are commonly expressed as 
$\sigma_0^2$, $\sigma_{01}$, and $\sigma_1^2$.  
The last component is then the residual variance $\sigma_r^2$.

\subsection{Marginal Expression}
Based on Equations~\ref{eq:lmmcond}, \ref{eq:lmmran},
and \ref{eq:lmmres}, the marginal distribution of the LMM is
\begin{equation}
  \label{eq:marginml}
  \bm y_j \sim N(\bm X_j \bm \beta, \bm V_j),
\end{equation}
where
\begin{equation}
  \label{eq:marginv}
    \bm V_j = \bm Z_j \bm D \bm Z_j^{\top} + \bm R_j, 
\end{equation}
where $\top$ denotes a matrix transpose.  By using the combined notation 
from Equation~\eqref{eq:redefine1} to 
Equation~\eqref{eq:redefine4}, we can further define $\bm V$ as
\begin{eqnarray}
  \label{eq:marginvtotal}
    \bm V &=& \bm V_{j} \otimes \bm I_{J}\\
          &=& \bm Z \bm G \bm Z^{\top} + \bm R.
\end{eqnarray}
Thus, Equation~\eqref{eq:marginml} can be rewritten as:
\begin{equation}
  \label{eq:marginml2}
  \bm y \sim N(\bm X \bm \beta, \bm V),
\end{equation}
From Equation~\eqref{eq:marginml2}, we can perceive the LMM as a 
regular linear model with correlated residual variance $\bm V$.  
From this perspective one can 
easily deduce that heterogeneity in $\bm V$ has little 
impact on the estimate of $\bm \beta$, because $\hat{\bm{\beta}}$ is still equal to 
$(\bm X^{T}\bm X)^{-1}\bm X \bm Y$, but can have a large impact 
on the significance test of $\bm \beta$ \cite{bates2015}.  We will illustrate 
this issue in the following section. 

\section{Problems Stemming from Heterogeneity}
We now illustrate implications of heterogeneity via both theoretical 
results and simulation.

\subsection{Theoretical Demonstration}
The variance-covariance matrix w.r.t. the fixed 
parameter corresponds to the inverse of the model's Fisher information, 
the relevant part of which can be expressed as \cite<e.g.,>{wanmer16}: 
\begin{eqnarray}
  \label{eq:vfix}       
  \bm V_{\beta} &=& (\bm X \bm V^{-1} \bm X^{T})^{-1}\\
  &=& (\bm X (\bm Z \bm G \bm Z^{T} + \bm R)^{-1}\bm X)^{-1}\\
  &=& \bm X^{-1}(\bm Z \bm G \bm Z^{T} + \bm R)(\bm X^{T})^{-1}.
\end{eqnarray}
The standard error of fixed parameters, $\text{SE}_{\bm{\beta}}$, 
is then the square root of the diagonal elements of $\bm V_{\bm{\beta}}$. 
This shows that $\bm V$ directly contributes to the fixed parameters' 
standard errors, which in turn influences the fixed parameters' test 
statistics. With the under/over-estimates of $\text{SE}_{\bm{\beta}}$, the 
\emph{t}-statistic will be larger/smaller than it should be. 
Generally one can expect that the increasing of 
$\bm V$ results in Type II error whereas decreasing of $\bm V$ leads to 
Type I error.  In practice, the former happens more often.  
\citeA{kwo07} conducted a series of Monte Carlo 
simulations and 
found underspecification and misspecification 
of $\bm V$ result in overestimation of $\text{SE}_{\beta}$, 
which lead to lower statistical power in significance tests of the 
fixed parameters.  Although their simulations only examined main 
effects, one can expect similar results for interaction 
effects. We illustrate this issue in the next section. 

\subsection{Data Demonstration}
In this section, we specifically illustrate how the change (increase) in 
$\bm V$ could impact the
significance of fixed parameters 
by using artificial data similar to the 
\emph{sleepstudy} data \cite{belenky03} 
included in \emph{lme4}. 
This dataset includes
$18$ subjects participating in a sleep deprivation study, where each
subject's reaction time (RT) \footnote{Strictly speaking, the response variable
should be log(RT) to have a meaningful infinity support.} 
was monitored for $10$ consecutive days.
The reaction times are nested within subject and 
continuous in measurement.  
Then we fit a model with day of measurement (``\emph{Days}'') as 
the covariate, including random
intercepts and slopes that are allowed to covary. 
This leads to
a model whose free parameters include: the fixed intercept and 
slope $\beta_0$ and $\beta_1$;
the random variance and covariances $\sigma_0^2$, $\sigma_1^2$,
and $\sigma_{01}$; and the residual variance $\sigma_r^2$.  To illustrate 
the impact of heterogeneity on cross-level interactions, we also simulate 
an ordinal variable with four levels loosely called 
\emph{Cognitive Ability} (CA), with its own main effect
coefficient as $\beta_2$ and its interaction effect 
\emph{Cognitive Ability} (CA) $\times$ \emph{Days} coefficient as $\beta_3$.
In the simulation, we focus on the significance test results of 
$\beta_1$ and $\beta_3$.  The true values were set to be $10.47$ and $6.27$, 
respectively, with both far different from $0$.  The random effect 
variance/covariance and residual variance were set to be the same as 
the estimates obtained from the \emph{sleepstudy} data.  
This leads to the model 
 displayed from Equation~\eqref{eq:prob1} to Equation~\eqref{eq:prob3}.
\begin{eqnarray}
  \label{eq:prob1}
  \text{RT}_j|\text{Subject}_j &\sim& N(\beta_0 + \beta_1\text{Days} 
  + \beta_2\text{CA}
  + \beta_3\text{Days} \times \text{CA}, \bm R_j)\\
  \label{eq:prob2}
  \text{Subject}_j &\sim& N(\bm 0, \left[ {\begin{array}{cc}
   \sigma_0^2 &  \sigma_{01}\\
   \sigma_{01} & \sigma_{1}^2\\
  \end{array} } \right]) \\
  \label{eq:prob3}
  \bm R_j &=& \sigma_r^2\bm I_{10}
\end{eqnarray}

From Equation~\eqref{eq:marginv}, it can be observed that 
changes in $\bm V$ can come from either $\bm G$, 
which is composed of between-subjects variance parameters
$\sigma_0, \sigma_1$ and 
$\sigma_{01}$, or the residual variance $\sigma_r^2$.  
We generated data so that  $\bm V$ changed with each of these 
four parameters, including the between 
subjects intercept variance $\sigma_{0}^2$, slope variance $\sigma_{1}^2$, 
covariance $\sigma_{01}$ and residual 
variance $\sigma_{r}^2$, along with different sample sizes as 
small ($n = 120$), medium ($n = 480$) and large ($n = 960$).  
Changes in these variance parameters began at 
cognitive ability level 2 and were consistent thereafter.  Participants below 
cognitive ability level 2 deviated
from participants at or above level 2 by $d$ times the 
parameters' asymptotic standard errors (scaled by $\sqrt{n}$), 
with $d = 0, 1, 2, 3, 4$.  To obtain the asymptotic 
standard error, we first fit a model under the above 
parameter settings but with a large sample size, e.g. $9600$; then 
the asymptotic standard error can be extracted by taking the square root of
the diagonal elements of the variance covariance matrix. 
The replication code for obtaining asymptotic standard error used 
throughout this paper is provided in the supplementary material.

\begin{figure}
\caption{\scriptsize{Simulated data sets without heterogeneity (left panel) 
and with heterogeneity (right panel).  The sample size for both examples is 120.   
}}
\label{fig:probdis}
\includegraphics{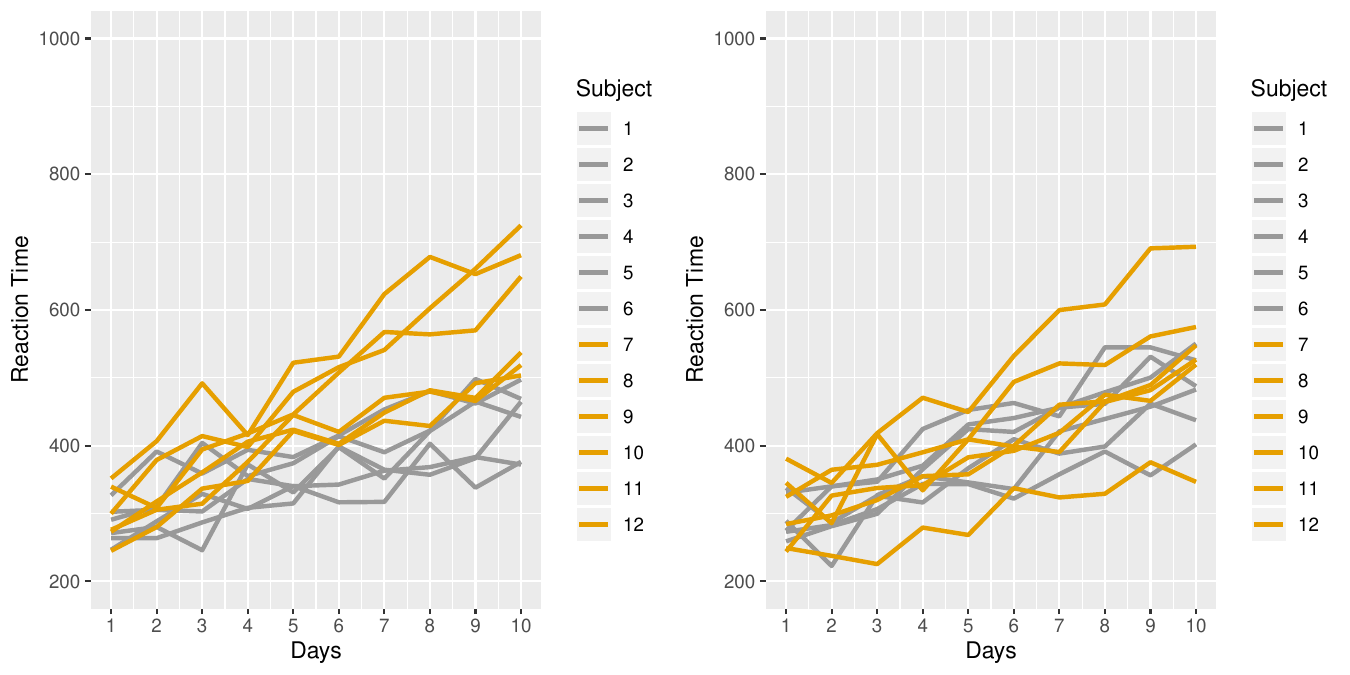}
\end{figure}

The magnitude of change is reflected in $d$.  When $d$ is 0, it 
represents homogeneity in the corresponding parameter, which serves as the 
baseline; when $d$ is greater than 0, it 
represents 
heterogeneity in $\bm V$ (increasing with Cognitive Ability in this example), 
with larger $d$ indicating more severe heterogeneity.  One example of data 
with and without heterogeneity is displayed in Figure~\ref{fig:probdis}.  In 
the left panel, data were generated without heterogeneity in random slope 
($d = 0$); whereas in the right panel data were generated with heterogeneity in 
random slope as large as $d = 4$.  Within each panel, 
the subjects 1-6 denoted with gray 
lines have cognitive ability equal to or less than 2, 
whereas subject 7-12 denoted with yellow lines 
have cognitive ability greater than 2.  Without the 
impact of variance heterogeneity (left panel), it is 
easy to observe that the RT has a positive relation 
with Days ($\beta_1$), and this relation
differ for subjects with different cognitive abilities $(\beta_3)$.  
However, these relations are
difficult to see under the impact of variance change (right panel).  
Unfortunately, the real 
data often look more similar to the right panel, with no obvious 
relations to be detected 
even if the generating fixed parameters are actually exactly 
the same as the left panel.  
To formally examine the impact of 
heterogeneity, we computed the percentage of 
significant fixed parameters related to ``Days'' 
($\beta_{1}$  and $\beta_3$) among $1,000$ 
replications in each condition.

\begin{figure}
\caption{\scriptsize{Simulated power curves for $\beta_{1}$ and 
    $\beta_3$ 
    under situations with heterogeneity in parameters 
    $\sigma_{r}^2$, $\sigma_{1}^2$, $\sigma_{0}^2$ and $\sigma_{01}$ 
    across ranges of 0--4 asymptotic variance. Panel labels denote 
    the parameter being tested and the parameter with heterogeneity.}}
\label{fig:probres}
\includegraphics{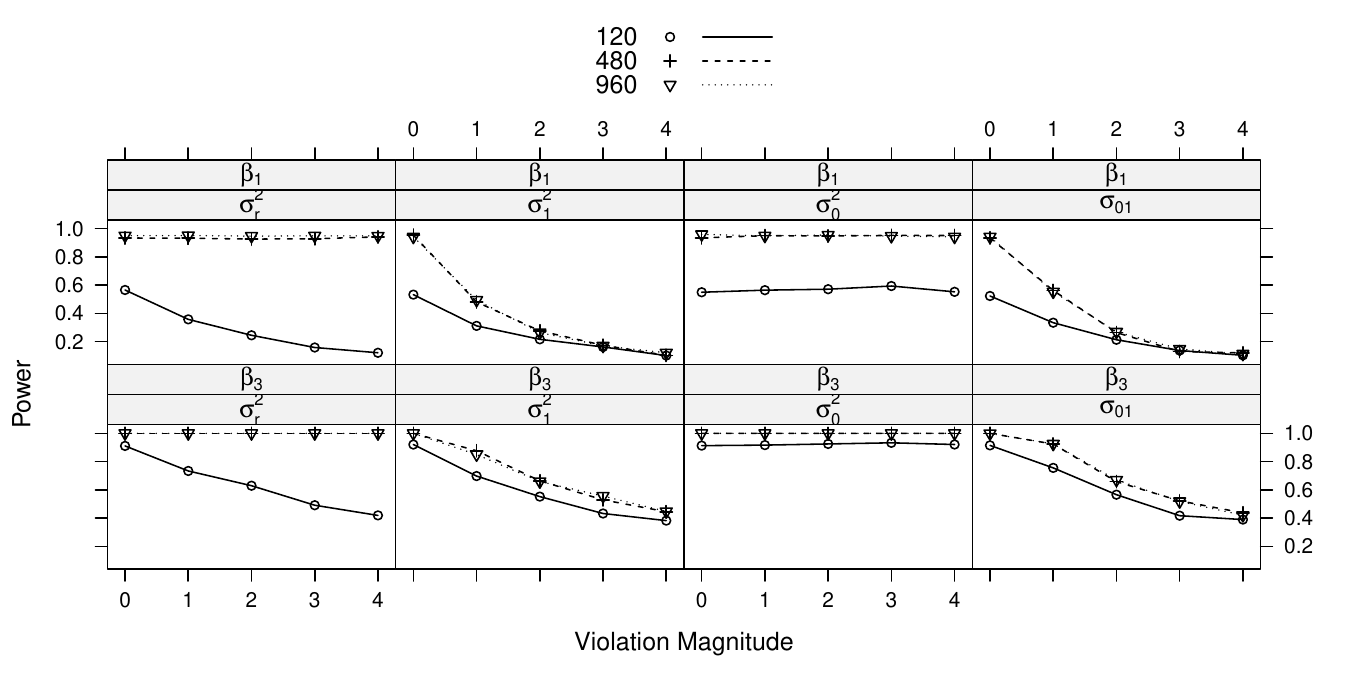}
\end{figure}

The full simulation results for $\beta_{1}$  and $\beta_3$ are demonstrated 
in Figure~\ref{fig:probres}, 
with the panel titles first indicating the tested parameter and
then indicating the heterogeneous parameter, and the y-axis representing power
(using $\alpha = 0.05$).  In general, when sample size is medium or 
large, increasing heterogeneity in the slope variance $\sigma_1^2$ or 
covariance $\sigma_{01}$ reduces power for both the main 
effect and interaction effect.  Heterogeneity  
in the residual variance or
intercept variance does not impact power for $\beta_1$ or $\beta_3$, 
because they can be compensated for during 
estimation \cite{kwo07}.  That is to say, when the intercept 
variance (or residual variance) increases, the residual 
variance (or intercept variance) estimate will decrease to 
compensate for the change, leading to the diagonal of $\bm V$ being 
unchanged.  This compensation effect exists because 
the intercept covariate in the random effect design matrix ($\bm Z$) is 
all $1$, so that the intercept and residual variance contribute equally to the diagonal of $\bm V$.  

When sample size is small ($n = 120$), power is generally lower 
in all scenarios.  In addition, greater heterogeneity in the residual 
variance also leads to lower power, which might be due to the fact that
heterogeneity combined with small sample size 
is more likely to result in unstable variance/covariance estimates, or even 
convergence issues. 

Overall, however, failing to account for the upward changes 
in $\bm V$ would generally 
result in Type II error.  
Although it is important to systematically monitor heterogeneity in
variance components,
it is also plausible that a fixed parameter indeed changes 
according to another variable (e.g., that an interaction exists).
Ideally, there would exist a statistical test that can
differentiate between these two kinds of changes.  
In the next section, we will introduce a score-based family of statistical 
tests that can fulfill this need.

\section{Score-based Tests}
In this section, we will introduce the score-based test as applied to the 
framework of LMM.  This introduction draws on LMM results described 
by \citeA{wanmer16} and is related to tests described 
by, e.g., \citeA{zeihor07}, \citeA{MerFanZei}, and \citeA{WanMerZei14}.

\subsection{Scores}
Based on the marginal model expression shown in Equation~\eqref{eq:marginml2}, 
the log likelihood of the LMM can be 
expressed as: 

\begin{equation}
  \label{eq:obj}
    \ell(\bm \sigma^2, \bm \beta; \bm y) = -\frac{n}{2}\log(2\pi) -
    \frac{1}{2}\log(|\bm V|) - \frac{1}{2}
    (\bm y - \bm X \bm \beta)^{\top}\bm V^{-1} (\bm y- \bm X \bm \beta).
\end{equation}
Scores, denoted $s_i()$ in this paper, are based on the first partial 
derivatives of $\ell$ w.r.t. $\bm \theta = (\bm \sigma^2\ \bm \beta)^\top$. 
The scores involve these partial derivatives evaluated for each observation 
$i$, where $i$ = ${1, 2, \ldots, n}$, 
and they can be roughly viewed as a residual: values close to 0 imply that the model 
provides a good fit to case $i$ with respect to a specific parameter, and values far from 0 imply the opposite.

The model gradient is equal to the sum of scores across all individuals 
and clusters:
\begin{eqnarray}
  \label{eq:firstder}
    \ell^{'}(\bm \theta; \bm y) &=& \sum_{j=1}^{J}
    \frac{\partial \ell(\bm \theta; \bm y_{j})}
    {\partial \bm \theta} = \sum_{j=1}^{J}
    \sum_{i \in c_j} s_i(\bm \theta; y_i)
\end{eqnarray}
where $\frac{\partial \ell(\bm \theta; \bm y_{j})}{\partial \bm \theta}$
represents the first derivative within cluster $c_j$, which can be 
expressed as the sum of the casewise score $s_i()$ belonging to 
cluster $j$.  For LMMs, the function $s_i()$ w.r.t. $\sigma_{k}^{2}$ 
and $\bm \beta$ can be expressed as the $i$th component of the 
vectors or as the $i$th row of the matrix (if $\bm \beta$ 
contains multiple components):

\begin{align}
  \label{eq:scoresigma}
    s(\sigma_k^2; \bm y) &= -\frac{1}{2}
    \text{diag} \left [\bm V^{-1} \frac{\partial \bm V}
      {\partial \sigma_k^2}\right ] +
    \left \{\frac{1}{2}(\bm y-\bm X \bm \beta)^{\top} \bm V^{-1} \left
    (\frac{\partial \bm V}{\partial \sigma_k^2}\right ) \bm V^{-1}\right\}^{T}
    \circ (\bm y- \bm X \bm \beta) \\
  \label{eq:scoresbeta}
    s(\bm \beta; \bm y) &= \left\{\bm X^{\top} \bm V^{-1} \right\}^{T} \circ (\bm y-\bm X
      \bm \beta),
\end{align}
where $\circ$ represents the Hadamard product (component-wise multiplication). 
Further detail on these derivations can be found in \citeA{wanmer16}, \citeA{mcc01}, and \citeA{stroup12}.

These equations provide scores for each observation $i$, and we can
construct the clusterwise scores by summing scores within each cluster.
In situations with one grouping (clustering) variable, the clusterwise 
scores can be obtained from a fitted \pkg{lme4} model via R 
package \pkg{merDeriv} \cite{merDeriv}. 

\subsection{Statistics}
As applied to the LMMs considered here, score-based tests can be used to 
study heterogeneity that is potentially explained by an auxiliary 
variable $T$; for example, in the data demonstration considered earlier, 
the auxiliary 
variable could have been Cognitive Ability. Because the scores can be viewed 
as a type of residual, the score-based tests basically help us judge 
whether the residual magnitudes are associated with $T$. Because we have 
unique scores for each model parameter, we can also obtain information 
about where heterogeneity occurs.

Statistically, the tests considered here can be viewed as  
generalizations of the 
Lagrange multiplier test. The tests are based on a cumulative sum 
of scores, where the order of accumulation is determined by $T$. If there 
is no heterogeneity explained by $T$, then this cumulative sum should 
fluctuate around zero. Otherwise, the cumulative sum would systematically 
diverge from zero.

To formalize these ideas, we first define the (scaled) cumulative sum of 
the ordered scores.  This can be written as
\begin{equation} 
    \label{eq:cumscore}
  {\bm B}(t; \hat {\bm \theta}) ~=~ \hat {\bm I}^{-1/2} J^{-1/2}
    \sum_{j = 1}^{\lfloor J \cdot t \rfloor} {\bm s}(\hat {\bm \theta}; y_{(j)})
  \qquad (0 \le t \le 1)
\end{equation}
where $\hat{\bm I}$ is an estimate of the information matrix, $\lfloor
jt \rfloor$ is 
the integer part of $jt$ (i.e., a floor 
operator), and $x_{(j)}$ reflects the cluster with the $j$-th
smallest value of the auxiliary variable $T$.  While the above
equation is written in general form, we can
restrict the value of $t$ in finite samples to the set $\{0,
1/J, 2/J, 3/J, \ldots, J/J\}$.
We focus on how the
cumulative sum fluctuates as more clusters' scores are added to it, e.g.,
starting with the person of lowest cognitive ability and ending with the 
person of highest cognitive ability.
The summation is premultiplied by an estimate of the inverse square
root of the information matrix, which serves to decorrelate the
fluctuation processes associated with model parameters.  For LMMs,  
$\hat{\bm I}$ can be written as expected information matrix
\cite<e.g.,>{wanmer16}: 

\begin{equation*}
        \hat{\bm I} = \left[\begin{array}{ccc|ccc}
                   &&\\
                   & \bm X^{\top} \bm V^{-1} \bm X &&& \bm 0 &\\
                   &&\\
                   \hline
                   &&\\
                   & \bm 0 &&& \bm \left (\frac{1}{2}\right) \text{tr}
      \left [ \bm V^{-1} \left (\frac{\partial \bm V}
      {\partial \sigma_{k1}^2}\right)
      \bm V^{-1} \left (\frac{\partial \bm V}
      {\partial \sigma_{k2}^2}\right)\right ]&\\
                   &&\\
                   \end{array}\right].
\end{equation*}

\subsection{Score-based tests statistics}
To obtain an official test statistic, we must summarize the behavior of 
the cumulative sum in a scalar. Multiple summaries are available, leading 
to multiple tests of the same hypothesis.  For example, 
one could take the absolute maximum that the cumulative sum attains for any
parameter of interest, resulting in a {\em double max} statistic 
(the maximum is taken across parameters and clusters entering into 
the cumulative sum).
Alternatively, one could sum the (squared) cumulative sum across parameters
of interest and take the maximum or the average across clusters, 
resulting in a 
{\em maximum Lagrange multiplier} statistic and \emph{Cram\'{e}r-von Mises} statistic, respectively 
\cite<see>[for further discussion]{merzei13}.  These statistics are given by
\begin{eqnarray}
    \label{eq:dmax}
    \mathit{DM} & = & \max_{j = 1,\dots, J} \max_{k = 1, \dots, K} | {\bm B}(\hat {\bm \theta})_{jk} | \\
        \label{eq:cvm}
    \mathit{CvM}     & = & J^{-1} \sum_{j = 1,\dots, J} \sum_{k = 1, \dots, K} {\bm B}(\hat {\bm \theta})_{jk}^2, \\
    \label{eq:maxlm}
    \max \mathit{LM} & = & \max_{j = \underline{j}, \dots, \overline{\jmath}} ~
      \left\{ \frac{j}{J} \left( 1 - \frac{j}{J} \right) \right\}^{-1}
      \sum_{k = 1, \dots, K} {\bm B}(\hat {\bm \theta})_{jk}^2.
\end{eqnarray}

For an ordinal auxiliary variable $T$ with $m$ levels, we can modify the 
statistics above so that the maximum is only considered after all 
clusters at the same level of $T$ have entered the summation. This leads 
to test statistics that are especially sensitive to heterogeneity that 
is monotonic with $T$ \cite{MerFanZei}.
Formally, we define $t_{L}$ $(L=1,\ldots,m-1)$ to be the empirical, cumulative 
proportions of clusters observed at the first $m-1$ levels of $T$.
The modified statistics are then given by
\begin{eqnarray}
    \label{eq:wdm}
      \mathit{WDM}_o & = & \max_{j \in \{j_1, \dots, j_{m-1} \}} ~ \left\{ \frac{j}{J} \left( 1 - \frac{j}{J} \right) \right\}^{-1/2}      
                             \max_{k = 1, \dots, K} | {\bm B}(\hat {\bm \theta})_{jk} |,\\
    \label{eq:maxlmo}
      \max \mathit{LM}_o & = & \max_{j \in \{j_1, \dots, j_{m-1} \}} ~
                             \left\{ \frac{j}{J} \left( 1 - \frac{j}{J} \right) \right\}^{-1}
                             \sum_{k = 1, \dots, K} {\bm B}(\hat {\bm \theta})_{jk}^2,
\end{eqnarray}
where $j_{L}=\lfloor n \cdot t_{L} \rfloor$ $(L=1,\ldots,m-1)$.

Finally, if the auxiliary variable $T$ is only nominal/categorical, the 
cumulative
sums of scores can be used to obtain a Lagrange multiplier statistic.
This test statistic can be formally written as
\begin{equation}
    \label{eq:lmuo}
     \mathit{LM}_{uo} = \sum_{L = 1, \dots, m} \sum_{k = 1, \dots, K}
       \left( {\bm B}(\hat {\bm \theta})_{j_L k} - {\bm B}(\hat {\bm \theta})_{j_{L - 1}k} \right)^2,
\end{equation}
where ${\bm B}(\hat {\bm \theta})_{j_{0}k} = 0$ for all $k$.
This statistic is asymptotically equivalent to the usual, likelihood
ratio test statistic, and it is advantageous over the likelihood ratio
test because it requires estimation of 
only one model (the restricted model).  We make use of this advantage in the
simulations, described later.

In the following section, we apply these tests to a linear mixed 
model with one grouping variable, studying the 
tests' ability to distinguish between heterogeneity and interactions.

\section{Simulation}
The goal of the simulation is to examine score-based tests' abilities to 
differentiate between changes in fixed effect 
parameters (i.e., interaction effects) and changes in variance 
parameters (i.e., heterogeneity). For ease of description, we frame the 
data-generating model as being based on a longitudinal 
depression intervention administered to participants with different 
levels of cognitive ability (here, we assume that $m = 4$, i.e., that there 
are four ordered levels of cognitive ability).  
Each participant's depression magnitude is measured 
once per month during a 10 month period.  Thus $10$ measurements 
are nested within each participant, which comprises a typical application 
for LMMs.  It is plausible that the amount of time needed to change the 
magnitude of depression is dependent on subjects' cognitive ability.  
If so, there exists an interaction 
between time and cognitive ability.  However, it is also possible that 
patients with higher cognitive ability have larger intercept 
variance ($\sigma_0^2$) or residual variance ($\sigma_r^2$).  In addition, 
the interaction and heterogeneity might occur simultaneously.  
Since both interaction and heterogeneity can be viewed as parameter 
instability w.r.t. an auxiliary variable, we aim to examine the extent to 
which the score-based tests could attribute the parameter instability to the 
truly changing parameter(s) in an LMM framework. 

\subsection{Method}
Data were generated from an LMM.  The predictor is time, with its 
associated coefficient as $\beta_1$, and $\beta_0$ serves as the 
fixed intercept, which completes the fixed parameters in the model.  
We have covarying intercept and slope random effects as well, with the 
variance and covariance as $\sigma_0^2$, 
$\sigma_{01}$, and $\sigma_1^2$.  The variance not captured by 
the random effects is modeled by the residual 
variance $\sigma_r^2$.  
The true parameter change can occur in one 
of seven ways: fixed intercept $\beta_0$,  
time coefficient $\beta_1$, random intercept variance $\sigma_0^2$,  
random covariance $\sigma_{01}$, random slope variance $\sigma_{1}$, 
residual variance $\sigma_r^2$, 
or $\beta_1$ and $\sigma_r^2$ simultaneously.  
The fitted models matched the data 
generating model, and parameter estimates were obtained by marginal maximum 
likelihood.  Parameter changes were tested in each of the $6$ 
estimated parameters, respectively. 

Power and Type I error were examined across three sample 
sizes (n = 120, 480, 960), five magnitudes of parameter change, and six 
tests of each individual parameter.  
The parameters' true values were set to be the same as the estimates from 
the \emph{sleepstudy} data included in \emph{lme4}.  The parameter 
change point and changing magnitude is manipulated in the same way as 
the problem demonstration simulation. 

For each combination of sample size ($n$) $\times$ 
violation magnitude ($d$),
$1,000$ data sets were generated and all parameters were tested. 
Two ordinal statistics ($\text{maxLM}_o, \text{WDM}_o$) and one categorical
statistic were examined ($\text{LM}_{uo}$) \cite{MerFanZei, WanMerZei14}.
The categorical statistic is asymptotically equivalent to the usual
likelihood ratio test.  Thus, this statistic provides information about the
relative performance of the ordinal statistics vs.\ the LRT.

\subsection{Results}
\begin{figure}[H]
\caption{\scriptsize{Simulated power curves for $\max \mathit{LM}_o$,
$\mathit{WDM}_o$, and $\mathit{LM}_{uo}$ across parameter change of
         0--4 asymptotic standard error. 
         The changing parameter is $\beta_{0}$. Panel labels denote
         the parameter being tested along with sample size.}}
\label{fig:sim11res}
\includegraphics{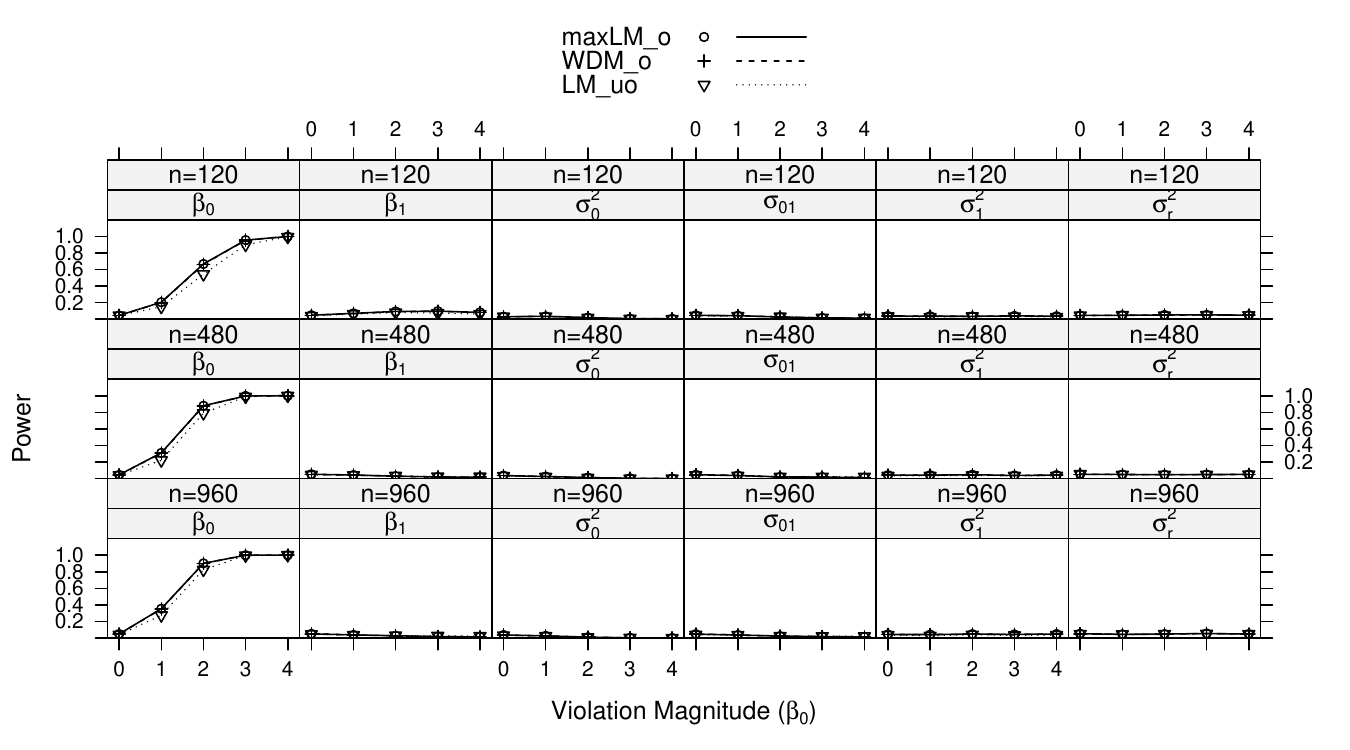}
\end{figure}

\begin{figure}[H]
\caption{\scriptsize{Simulated power curves for $\max \mathit{LM}_o$,
$\mathit{WDM}_o$, and $\mathit{LM}_{uo}$ across parameter change of
         0--4 asymptotic standard error. 
         The changing parameter is $\beta_{1}$. Panel labels denote
         the parameter being tested along with sample size.}}
\label{fig:sim12res}
\includegraphics{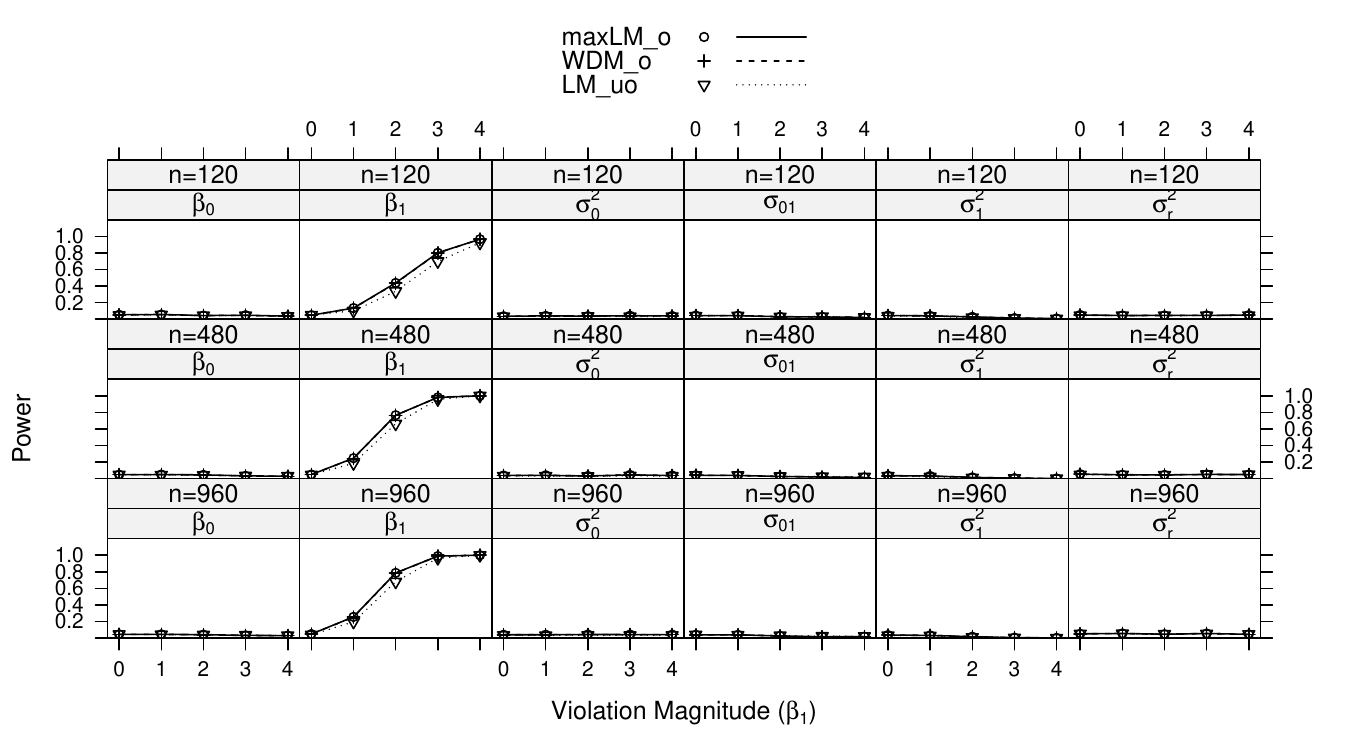}
\end{figure}

\begin{figure}[H]
\caption{\scriptsize{Simulated power curves for $\max \mathit{LM}_o$,
$\mathit{WDM}_o$, and $\mathit{LM}_{uo}$ across parameter change of
         0--4 asymptotic standard error. 
         The changing parameter is $\sigma_{0}^2$. Panel labels denote
         the parameter being tested along with sample size.}}
\label{fig:sim13res}
\includegraphics{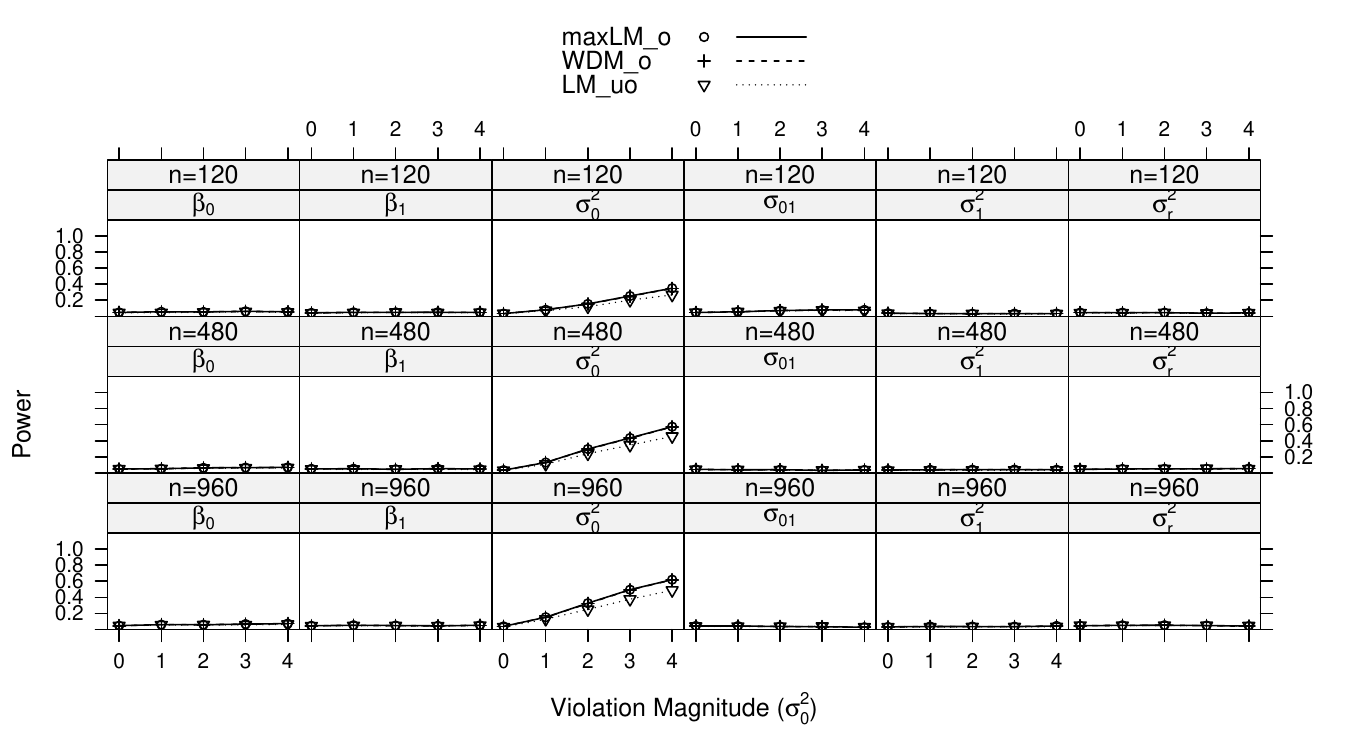}
\end{figure}

\begin{figure}[H]
\caption{\scriptsize{Simulated power curves for $\max \mathit{LM}_o$,
$\mathit{WDM}_o$, and $\mathit{LM}_{uo}$ across parameter change of
         0--4 asymptotic standard error. 
         The changing parameter is $\sigma_{01}$. Panel labels denote
         the parameter being tested along with sample size.}}
\label{fig:sim14res}
\includegraphics{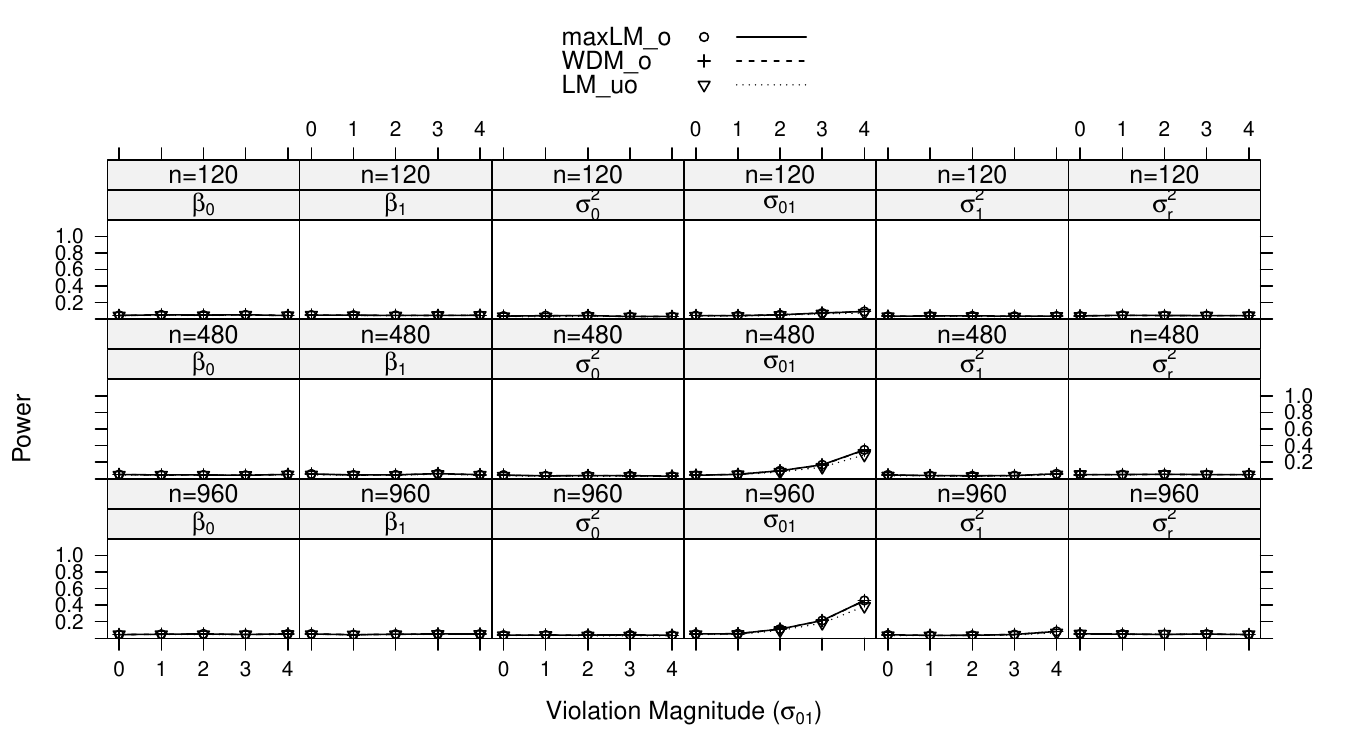}
\end{figure}

\begin{figure}[H]
\caption{\scriptsize{Simulated power curves for $\max \mathit{LM}_o$,
$\mathit{WDM}_o$, and $\mathit{LM}_{uo}$ across parameter change of
         0--4 asymptotic standard error. 
         The changing parameter is $\sigma_{1}^2$. Panel labels denote
         the parameter being tested along with sample size.}}
\label{fig:sim15res}
\includegraphics{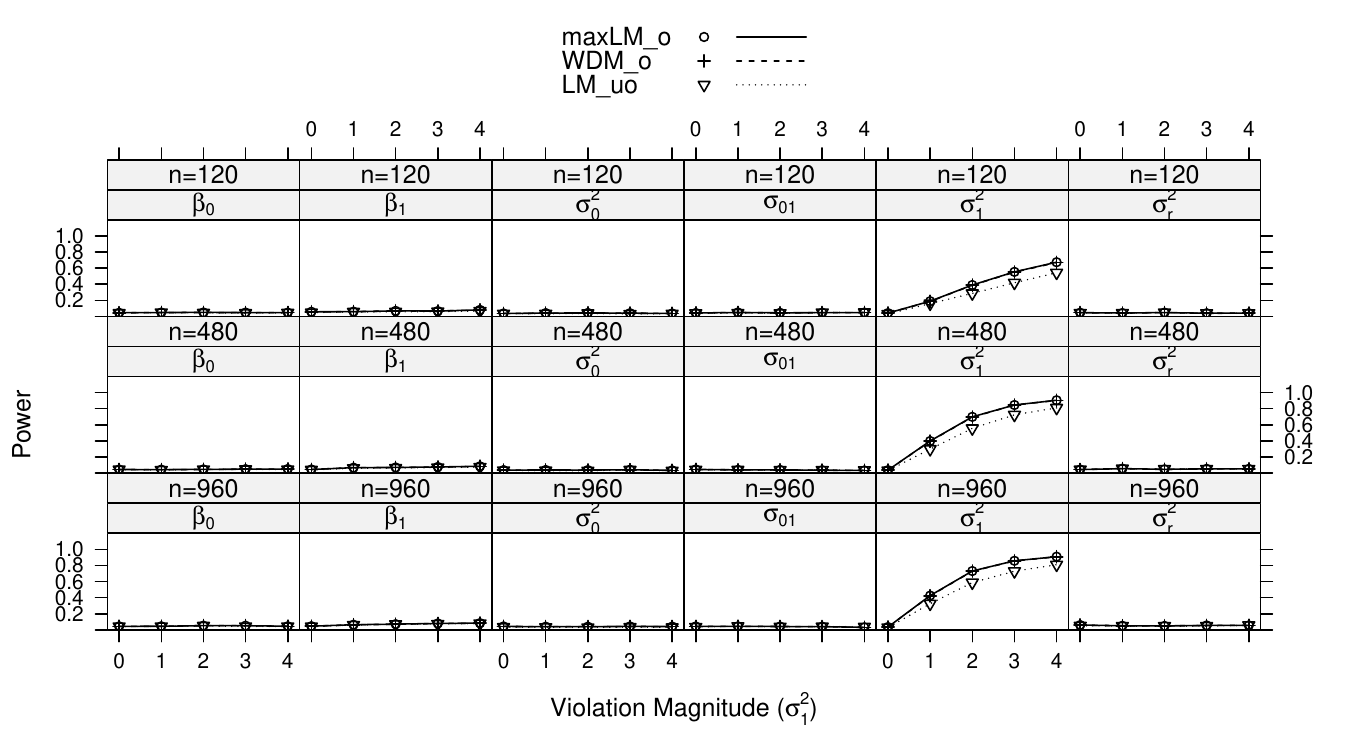}
\end{figure}

\begin{figure}[H]
\caption{\scriptsize{Simulated power curves for $\max \mathit{LM}_o$,
$\mathit{WDM}_o$, and $\mathit{LM}_{uo}$ across parameter change of
         0--4 asymptotic standard error. 
         The changing parameter is $\sigma_{r}^2$. Panel labels denote
         the parameter being tested along with sample size.}}
\label{fig:sim16res}
\includegraphics{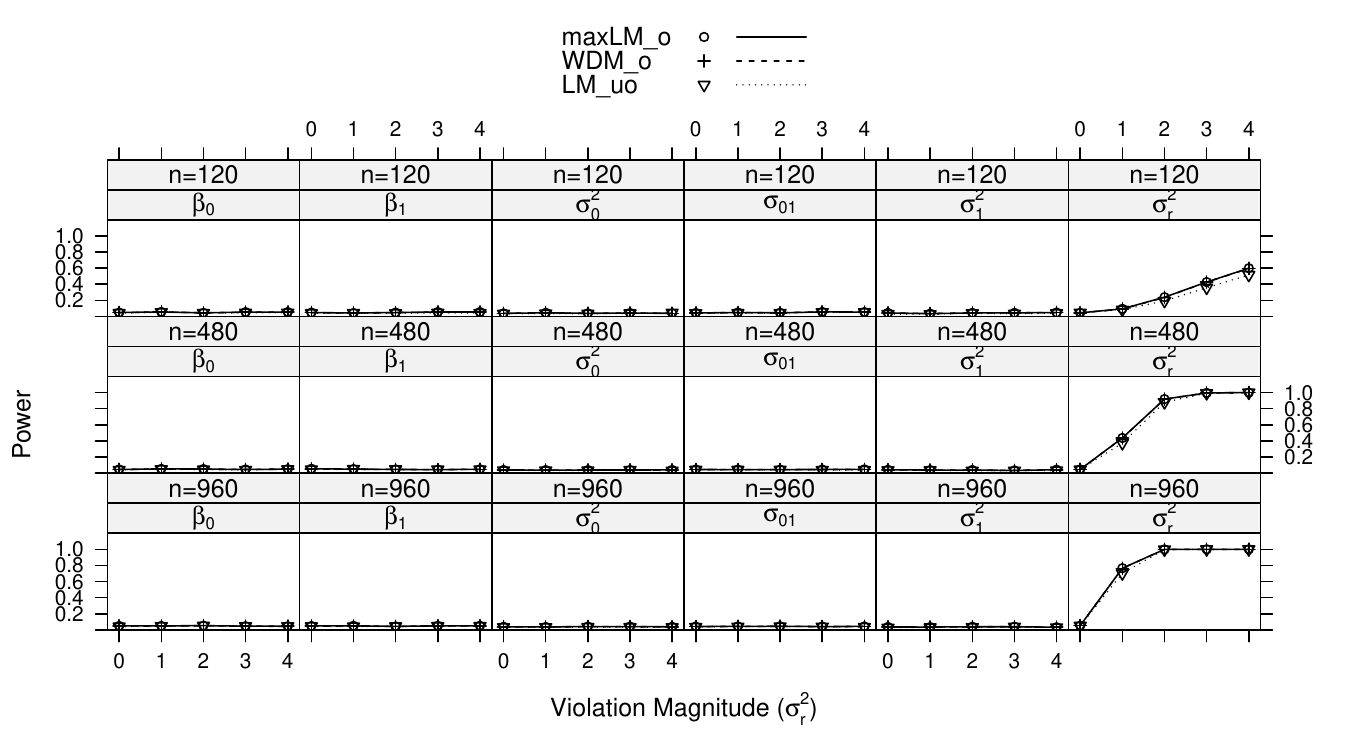}
\end{figure}

\begin{figure}[H]
\caption{\scriptsize{Simulated power curves for $\max \mathit{LM}_o$,
$\mathit{WDM}_o$, and $\mathit{LM}_{uo}$ across parameter change of
         0--4 asymptotic standard error. 
         The changing parameter are $\beta_{1}$ and $\sigma_{r}^2$. Panel labels denote
         the parameter being tested along with sample size.}}
\label{fig:sim17res}
\includegraphics{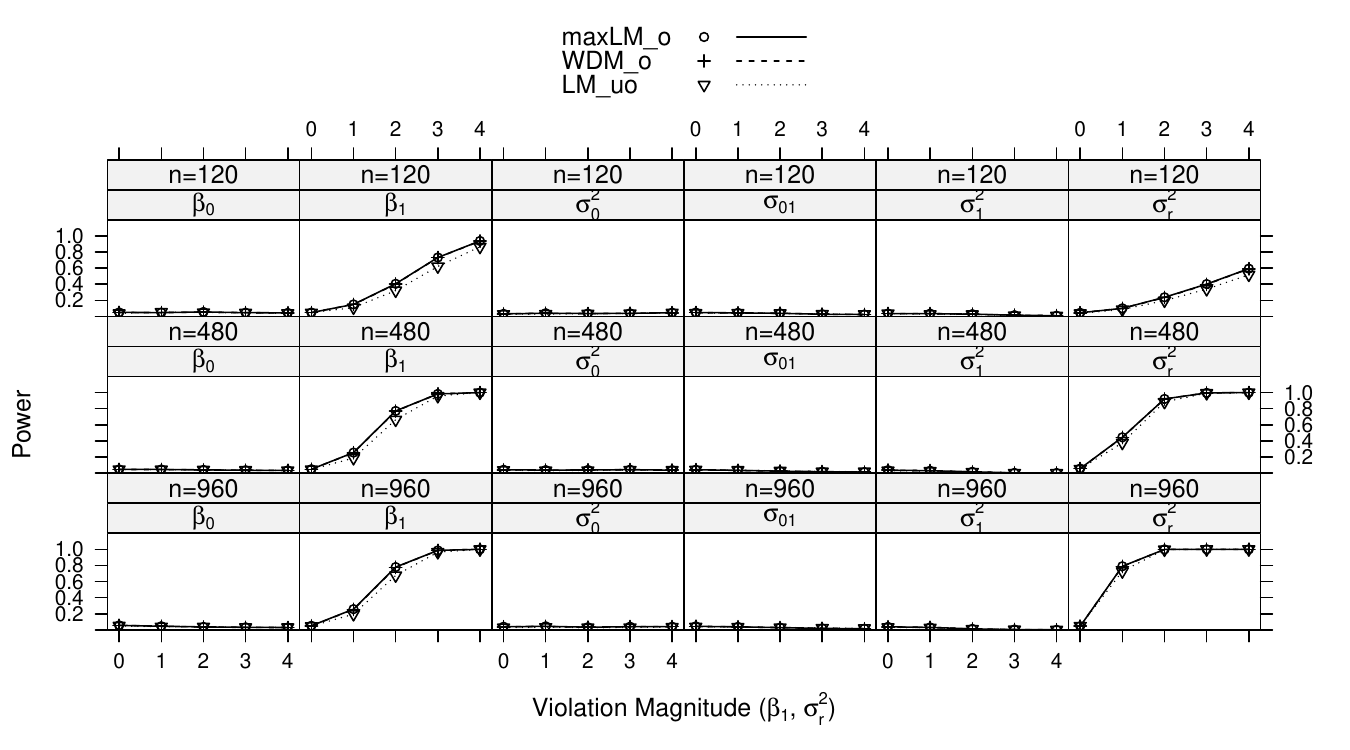}
\end{figure}

Full simulation results are presented in 
Figures~\ref{fig:sim11res} to~\ref{fig:sim17res}. 
In each graph, the x-axis represents
the violation magnitude and the y-axis represents the 
power of detecting parameter change.  When $x = 0$, the corresponding 
power serves as the type I error. 
Figure~\ref{fig:sim11res} demonstrates power curves as a function of
violation magnitude in $\beta_0$, with sample 
size changing 
across rows, the tested parameters changing
across columns,  and lines reflecting different test statistics.
Figure~\ref{fig:sim12res} to 
Figure~\ref{fig:sim16res} 
display similar power curves when the true changing parameter 
is $\beta_1$, 
$\sigma_0$, $\sigma_{01}$, $\sigma_{1}$ and 
$\sigma_r^2$, respectively.  
Figure~\ref{fig:sim17res} shows the power 
curves when there exist two changing parameters, $\beta_1$ and $\sigma_r^2$.

From these figures, one can generally 
observe that the score-based statistics could 
isolate the truly-changing parameter, with non-zero power curves for 
changing parameter(s), and near-zero power curves for non-changing 
parameters.  For example, in Figure~\ref{fig:sim11res}, 
for $\beta_0$, 
the power increases with the violation magnitude $d$ and sample 
size (across rows); the power for the other five non-changing 
parameters remain near zero (across columns), even with increasing 
violation magnitude and sample size.  

Within each non-zero power curve panel of 
Figure~\ref{fig:sim11res} to 
Figure~\ref{fig:sim17res}, 
the two ordinal statistics, $\text{maxLM}_o$ and $\text{WDM}_o$, exhibit higher
(when testing fixed parameter or random intercept variance) or 
similar (when testing residual variance) power compared with categorical 
statistic $\text{LM}_{uo}$.  This is partially consistent with the 
results demonstrated in \citeA{MerFanZei}, where ordinal statistics are shown 
to be more sensitive to monotonic parameter changes. The  
residual variance results (Figure~\ref{fig:sim16res}) might be due to a 
ceiling effect, where all three power curves quickly increase to 1.  
In conditions with only one changing parameter (Figure~\ref{fig:sim11res} to 
Figure~\ref{fig:sim16res}), 
$\text{maxLM}_o$ and $\text{WDM}_o$ are 
mathematically equivalent \cite{MerFanZei}.  
In conditions with 
two changing parameters 
(Figure~\ref{fig:sim17res}), $\text{maxLM}_o$ 
and $\text{WDM}_o$ still demonstrate similar power curves. 
The advantages of $\text{WDM}_o$ are only apparent when testing 
many (more than two) parameters at a time \cite{MerFanZei,WanMerZei14}.  

Comparing the non-zero power curves across these seven 
figures, it shows the score-based tests have somewhat higher power to 
detect residual 
variance change when sample size is medium or large, followed by 
fixed parameter change 
and random variance/covariance parameter change.
This phenomenon is most apparent by comparing
Figure~\ref{fig:sim13res} to Figure~\ref{fig:sim15res}
with Figure~\ref{fig:sim17res}, 
with the power curve for the residual variance and fixed parameter  
approaching 1 in conditions with medium or large sample sizes, while the 
power curve for the random variance/covariance ranges from 0.4 
to 0.8 even for the greatest $d$ under large sample size.
The general difficulty to detect parameter changes in the $\bm G$ 
matrix is related to the fact that large parameter
changes in variance/covariance components often render $\bm G$ as 
numerically non-positive definite, resulting in 
correlations of 1 or model non-convergence.  The Discussion 
section provides more details on this issue.

In summary, we found that the score-based tests can attribute heterogeneity 
to the truly problematic parameter(s) in an LMM context. Additionally, the 
tests were more sensitive to changes in fixed effect parameters, as compared 
to variance parameters.  In the next section, we apply the tests to real data 
to illustrate the potential usage of score-based tests in practice. 
The general approach is to fit a 
LMM of interest, then obtain each parameter's 
score-based test statistics w.r.t. an 
auxiliary variable in level 2.  
If the variance (either random effect or residual) component is detected to 
have parameter instability, it indicates heterogeneity present in the data; 
if the fixed parameter demonstrates instability, then we can claim 
interaction between the covariate and the auxiliary variable.

\section{Tutorial}

In this section, we demonstrate how the score-based 
tests can be carried out in \proglang{R}, using package 
\pkg{lme4} \cite{lme4} for model
fit and \pkg{strucchange} \cite{ZeiLei02,Zei06} for testing, with the 
score computations handled in the background by 
\pkg{merDeriv} \cite{merDeriv}.  We use data from the 
1982 ``High School and Beyond'' survey funded by the National Center 
for Education Statistics (NCES), which is available in 
\proglang{R} package \pkg{mlmRev}.  In the dataset, 
7185 U.S. high-school students from 
160 schools completed a
math achievement test, with the students' socioeconomic 
status (\code{ses}) as a level 1 covariate.  We center the \code{ses} covariate by 
school mean and focus on the centered \code{ses} (denoted as \code{cses}) 
below, as recommended by 
previous researches \cite{alg11, bau05, end07}. The centering only 
eases parameter interpretation, and generally has no impact on the 
cross-level interaction term's statistical significance test \cite{alg11}.

The aim of the current analyses is to determine how students' math 
achievement scores (denoted as \code{mAch} in the dataset) are associated 
with their family socioeconomic status.
It is plausible that the relationship between \code{cses} and 
math achievement differs across schools with different \code{meanses} 
(level 2 covariate).  The traditional approach is to fit the linear 
mixed model with an interaction term and examine the significance of 
the coefficient for the interaction term.
  
\subsection{Traditional Approach}
The traditional approach to testing the interaction between 
  \code{cses} and \code{meanses} can be carried out via

\begin{Schunk}
\begin{Sinput}
> library("mlmRev") 
> library("lme4")  
> library("lmerTest") 
> m2 <- lmer(mAch ~ cses*meanses + (cses|school), data = Hsb82, REML = FALSE)
> summary(m2)
\end{Sinput}
\end{Schunk}
where \code{cses*meanses} specifies the model fixed 
effects (both main effects and 
interaction), and \code{(cses|schools)} specifies the random 
effects; \code{REML = FALSE} 
requests the marginal maximum likelihood estimation as described 
in Equation~\eqref{eq:obj}.  
From the results returned by \code{summary()} (not shown), we can 
see the coefficient for the interaction term 
is not significant ($p = 0.367$).  However, as shown in the 
Figure~\ref{fig:probres}, 
the significance test for the interaction 
might be impacted by variance/covariance heterogeneity in random 
effects (second row of Figure~\ref{fig:probres}, second panel and 
fourth panel). 
Thus, we use the score-based tests to distinguish between the 
cross-level interaction and variance heterogeneity.

To conduct the score-based tests, we first fit the model with 
only the level 1 covariate, as shown in the code section below.  One advantage of 
score-based test is that the focal level 2 covariate does not enter the model but 
serves as the auxiliary variable in the testing stage.  This feature reduces 
model complexity and is more likely to lead to converged models.
\begin{Schunk}
\begin{Sinput}
> library("mlmRev") 
> library("lme4") 
> m1 <- lmer(mAch ~ cses + (cses | school), data = Hsb82, REML = FALSE)
\end{Sinput}
\end{Schunk}
This fitted model includes 6 parameters (with labels 1--6 below), 
which are $\beta_0$, $\beta_1$, $\sigma_0^2$ $\sigma_{01}$, $\sigma_1^2$ and 
$\sigma_r^2$ (in this order).  While the second level covariate \code{meanses}
would generally be treated as continuous, for demonstration purposes we 
consider treating it as continuous, ordinal, and categorical.  Each of these 
treatments is described separately below.

\subsection{Continuous Treatment}
If we treat the auxiliary variable \code{meanses} as 
continuous, we can employ \code{sctest()} to obtain continuous statistics 
from Equation~\eqref{eq:dmax} to 
Equation~\eqref{eq:maxlm} for the parameter of interest, which is 
specified by the \code{parm} argument. 

Because \code{sctest()} utilizes \code{estfun()} in the background, 
we need to ensure that the ordering of the auxiliary variable \code{meanses}
matches the row ordering of the 
\code{estfun()} output, so that each value of \code{meanses} corresponds to 
its associated \code{school}. The ordering and checking can be completed by 
the code below. 
This step is highly recommended in practice; the \pkg{data.table} 
package \cite{dat19} is utilized here solely for speed purposes.

\begin{Schunk}
\begin{Sinput}
> ## data.table for speed. 
> library("data.table")
> Hsb82 <- as.data.table(Hsb82)
> ## ordering 
> orderHsb82 <- Hsb82[order(as.numeric(school))]
> ## checking
> all.equal(rownames(estfun(m1)), unique(levels(orderHsb82$school)))
\end{Sinput}
\begin{Soutput}
[1] TRUE
\end{Soutput}
\end{Schunk}

After the ordering checking, we can proceed to the 
statistical tests.  For example, we can test whether  
the random slope variance ($\sigma_1$, specified
by \code{parm = 5}.) is stable across \code{meanses}.  
The code below displays how to conduct the tests.

\begin{Schunk}
\begin{Sinput}
> library("strucchange")  
> library("merDeriv")  
> dm <- sctest(m1, order.by = unique(orderHsb82$meanses), parm = 5, 
+   functional = "DM")
> cvm <- sctest(m1, order.by = unique(orderHsb82$meanses), parm = 5, 
+   functional = "CvM")
> maxlm <- sctest(m1, order.by = unique(orderHsb82$meanses), parm = 5, 
+   functional = "maxLM")
\end{Sinput}
\end{Schunk}
\begin{Schunk}
\begin{Sinput}
> c(dm$p.value, cvm$p.value, maxlm$p.value)
\end{Sinput}
\begin{Soutput}
[1] 0.013039947 0.009432531 0.002934033
\end{Soutput}
\end{Schunk}
We can see that all three statistics indicate significant 
parameter instability for the random slope variance parameter, 
suggesting the existence of heterogeneity.

\subsection{Ordinal Treatment}
In some scenarios, the school-level \code{ses} variable 
may only be measurable as ordered categories. To mimic this situation, 
we categorize schools with similar \code{meanses} to yield an 
ordinal level 2 auxiliary variable 
with five categories.  The code below first creates an ordinal variable 
and then shows that the only change in the \code{sctest()} command is 
the \code{functional} argument.

\begin{Schunk}
\begin{Sinput}
> # create ordinal variable
> orderHsb82[meanses < -0.5, schses:=1]
> orderHsb82[meanses <= -0.1 & meanses > -0.5, schses:=2]
> orderHsb82[meanses <= 0.1 & meanses > -0.1, schses:=3]
> orderHsb82[meanses <= 0.45 & meanses > 0.1, schses:=4]
> orderHsb82[meanses >= 0.45, schses:=5]
> # compute test statistics
> ordervar <- orderHsb82[,mean(as.numeric(schses)), by=school]$V1
> wdm <- sctest(m1, order.by = ordervar, parm = 5, 
+   functional = "WDMo")
> maxlmo <- sctest(m1, order.by = ordervar, parm = 5, 
+   functional = "maxLMo")
> c(wdm$p.value, maxlmo$p.value)
\end{Sinput}
\end{Schunk}
\begin{Schunk}
\begin{Soutput}
[1] 0.02977961 0.03029124
\end{Soutput}
\end{Schunk}
Like the previous statistics, both statistics are significant here as well.

\subsection{Categorical Treatment}
Lastly, when there is no ordering 
information contained in the auxiliary variable, categorical 
statistic can be implemented in the following way.  
This statistic is asymptotically equivalent to the traditional LRT as 
stated before, but has less power of detecting change as 
demonstrated in the simulation.  In this example, the test 
result is not significant ($\alpha = 0.05$) because the ordering information 
was ignored.
\begin{Schunk}
\begin{Sinput}
>  lmuo <- sctest(m1, order.by = ordervar, parm = 5, 
+    functional = "LMuo")
\end{Sinput}
\end{Schunk}
\begin{Schunk}
\begin{Sinput}
> lmuo$p.value
\end{Sinput}
\begin{Soutput}
[1] 0.06175341
\end{Soutput}
\end{Schunk}

\subsection{Subgroup Information}
\begin{figure}
  \caption{Empirical fluctuation processes of the $\max \mathit{LM}_o$
    statistic (first column) and $\mathit{WDM}_o$ (second column) for
    $\beta_0$ (first row), $\beta_1$
    (second row), $\sigma_0^2$ (third row), $\sigma_{01}$ (fourth row), 
    $\sigma_{1}^2$ (fifth row) and $\sigma_r^2$ (sixth row), 
    using M1 model. The statistics corresponding to the 5th
  numeracy level within each panel always equal to 0, 
  so not shown in the panels. }
 \label{fig:ordres}
 \center
 \includegraphics[height=8in, width=6in]{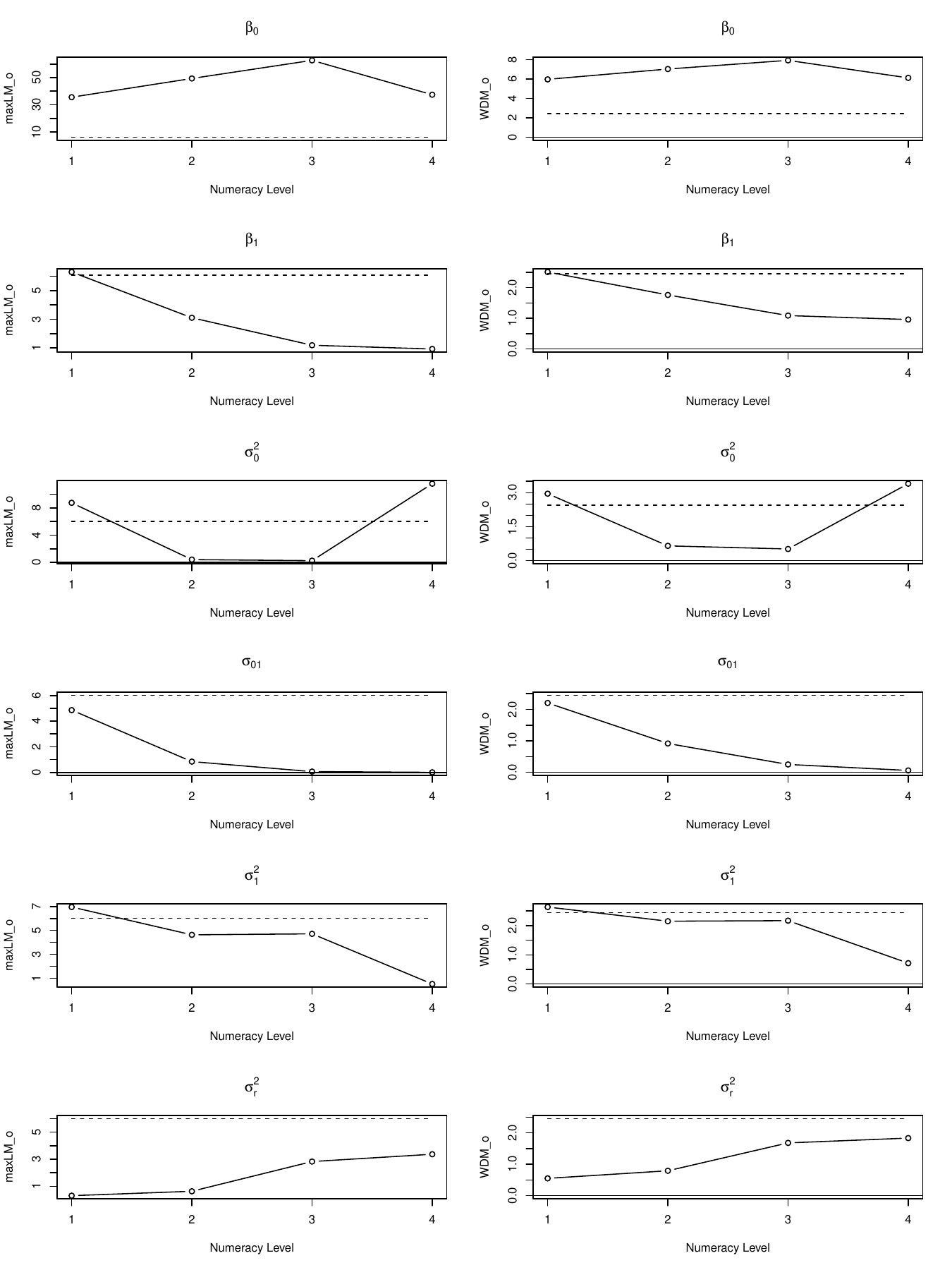}
\end{figure}

In addition to test statistics and $p$ value, 
``instability plots'' can be generated by setting \code{plot = TRUE} 
in the \code{sctest()} functions above.  Figure~\ref{fig:ordres} 
displays the ordinal statistics' fluctuation across \code{schses} levels.  
In this figure, the first column displays the fluctuation process associated 
with $\max \mathit{LM}_o$, and the second column 
displays the fluctuation process associated with $\mathit{WDM}_o$.  
Each panel represents the test of a specific model parameter, shown in 
the panel title.
Within each panel, the horizontal dashed line represents 
the 5\% critical value.  If the solid line crosses the critical value, then 
there is evidence that the corresponding parameter fluctuates across 
schses (because the full set of scores sum to zero, the final level of \code{schses} is 
not displayed on the x-axis).

In Figure~\ref{fig:ordres}, it is observed that 
the $\beta_0$, $\beta_1$, $\sigma_0^2$ and $\sigma_{1}^2$ demonstrate 
parameter instability, whereas 
$\sigma_{01}$ and $\sigma_r^2$ do not.  The instability of $\beta_0$ 
indicates that there exists a main effect of \code{schses}, and 
the instability of $\beta_1$ implies that there exists a cross-level
interaction effect between \code{schses} and \code{cses}.  In addition, 
the random intercept and the random slope demonstrate instability.  
As described earlier, this heterogeneity in random effect variances 
appears to have ``masked'' the significance of the interaction term.

Figure~\ref{fig:ordres} also provides 
information about levels of \code{schses} where parameters differ from one 
another; this can be discerned from levels where the solid line 
crosses the dashed horizontal line.
Thus, the intercept parameter changes w.r.t. each of the four levels of 
\code{schses} (all points are above the line); the slope of \code{ses} 
($\beta_1$) at \code{schses} level 1 differs from the other levels; 
the random intercept variance $\sigma_0^2$ has two changing points: one at 
level 1 and 
the other at level 4; and the random slope $\sigma_1^2$ differs between 
level 1 and the remaining levels. These results provide more detailed 
information about how \code{schses} is associated with different pieces 
of the model. 

To expand on the results above, we can create a dummy variable for students at 
\code{schses} level 1 (coded as 0), then use that dummy variable in place  
of \code{meanses}.  The code is given below:
\begin{Schunk}
\begin{Sinput}
> Hsb82[schses==1, dummy := 0]
> Hsb82[schses!=1, dummy := 1]
> m3 <- lmer(mAch ~ cses*dummy + (cses | school), data = Hsb82, REML = FALSE)
\end{Sinput}
\end{Schunk}

The interaction between cses and 
dummy variable is  1.09 with $p < 0.05$, 
indicating a significant interaction.  Thus, informed by the instability plots,
we can detect the ``masked''
interaction effect via traditional methods.  Alternatively, we can fit the model
separately for students at schools with the lowest level of \code{schses} and 
and for students at other schools with larger values of \code{schses}. The former model results in 
$\beta_1$ (coefficient of \code{cses}) 
as 1.21, whereas
the latter model has $\beta_1$ as 2.31.  
These results indicate that 
students' \code{cses} has stronger relationship with math achievement in 
schools with higher SES. 

In summary, score-based tests provide a statistical tool to 
closely examine an LMM's parameter estimates with respect to an 
auxiliary, level-2 variable.  The examination of variance 
components (random effect 
(co)variance and residual variance) provide tests of 
heterogeneity.  Additionally, the fluctuation plots can be used to 
interpret the nature of heterogeneity or interactions, without 
arbitrary median splits or subsamples of data with few observations. 

\section{Discussion}
In this paper, we extended a family of score-based tests to 
linear mixed models, focusing on models with one grouping variable. 
We found that the tests
can isolate specific parameters that exhibit instability, 
which avoids masked cross-level interaction effects in the presence 
of heterogeneity. They also provide specific information about groupings 
of the auxiliary variable whose parameter values differ. The tests 
developed in this paper can currently only be carried out on an auxiliary 
variable measured at the model's upper level (level 2), a 
restriction that leads to the future directions described below.

\subsection{Grouping with multiple variables}
The auxiliary variable is specifically required to be at the upper 
level because the tests described here require that the scores be 
independent. This independence assumption challenges models with at least 
two variables defining clusters, such as models with (partially) 
crossed random effects, or models with multilevel nested 
designs \cite<e.g.,>[Ch.\ 2]{bates10}.
In these cases, we cannot simply sum 
scores within a cluster to obtain independent, clusterwise scores, 
because observations in different clusters on the first 
grouping variable may be in the same cluster on the second grouping 
variable. 
A related issue occurs when the auxiliary variable is at the 
lowest (first) level of the model: scores at the lowest level are not 
independent, so the tests described here cannot be immediately used 
to test parameters with respect to a level-one variable. 

A natural approach to deal with the issue of dependent scores is to find a
heteroskedasticity and autocorrelation consistent (HAC)
covariance matrix estimator.  The traditional Hessian matrix only accounts for 
the correlations among score columns, whereas the HAC estimator is a robust 
Hessian estimator that can serve the purpose of de-correlating the scores 
under a generalized linear model framework.  
Several methods have been proposed here, including
kernel HAC estimators with
automatic bandwidth selection and weighted empirical
adaptive variance estimators \cite{and93}.  For multilevel
models with multiple grouping variables, \citeA{ber17} recently implemented 
a sandwich approach to obtaining robust variance/covariance matrices. 
It might be possible to deploy these methods in the context of linear 
mixed models. 

In practice, however, the technical challenge is to find the optimal bandwidth
in empirical studies with non-independent observations.  
Suboptimal selection of the bandwidth parameter could 
lead to decreasing power of detecting parameter change or even drop to 
zero \cite{per06}.  
\citeA{shao10} recently proposed use of a ``self-normalizing''
approach to tackle this 
technical issue, and they have utilized this approach in multivariate 
settings \cite{zhang11}.  An 
extension to linear mixed models with non-independent level 2 
grouping variables and non-diagonal residual covariance matrices is currently 
under development. 

\subsection{Model Estimation}
Along with independence issues, general model estimation issues 
may influence the score-based tests' accuracies. For example, in the 
relatively-common case where a parameter estimate lies on the 
boundary (e.g., a correlation between random effects near $\pm1$ or a 
variance approaching zero), then it may be impossible to carry 
out the proposed tests due to the non-positive definite structure 
of the model information matrix. 
Additionally, model misspecification can also influence the tests.  
One common type of misspecification involves the residual covariance matrix
having nonzero, off-diagonal elements that are fixed to zero in the estimated model. 
\citeA{WanMerZei14} examined the tests' 
performance in the factor analysis framework, and they found that unmodeled 
parameters' instability would lead the tests to identify instability 
in related model parameters.  In the same manner, we speculate that instability 
in unmodeled, off-diagonal residual covariances would be 
incorrectly attributed to random effects' variances or 
covariances ($\bm G$ components). 
Thus, it is important to carefully consider 
the specification of the estimated model.

\subsection{Tests' Power}
The power curves demonstrated in the simulation section are related to  
parameters' asymptotic standard errors and to sample size. In the simulation, 
we used the asymptotic variance-covariance matrix from the 
\code{sleepstudy} data, scaling these asymptotic standard error by square root
of the sample size.  However, as demonstrated in the tutorial section, 
applied researchers do not need to obtain the asymptotic standard 
error to utilize score-based tests.
Further, as demonstrated in the 
simulation, the power generally increases with sample size.  
The simulation sample sizes of
$120, 480, 960$ were used to conveniently allocate 
observations in 4 levels.  To achieve high power, there is a 
trade-off between sample size needed and the magnitude of instability.  
In the simulation, the small sample size of $120$ was sufficient 
because the instability was large enough.

\subsection{Summary}
In this paper, we generalized a family of score-based tests to 
two-level linear mixed models, which allow researchers to test whether 
model parameters fluctuate with an unmodeled level two variable. 
We found that the tests could successfully decouple cross-level 
interactions from variance heterogeneity, whereas heterogeneity 
could cause the traditional significance test of a cross-level 
interaction to exhibit inflated Type II error.  
Along with providing information about parameter stability across all 
estimated LMM parameters, the tests provide additional information 
about heterogeneous subgroups when parameter instability is detected. 
Thus, applied researchers in psychology and education can use the tests to 
examine potential cross-level interactions while ruling out possible 
masked results due to heterogeneity. 

\section*{Computational Details}
All results were obtained using the \proglang{R}~system for statistical
computing \cite{R11}, version~3.6.1,
employing the add-on package \pkg{lme4}~
1.1-21
\cite{lme4} for fitting of the linear mixed models and
\pkg{strucchange}~1.5-2
\cite{ZeiLei02,Zei06} for evaluating the parameter instability tests.
\proglang{R}~and both packages are freely available under the General
Public License from the Comprehensive \proglang{R} Archive Network at
\url{http://CRAN.R-project.org/}.
\proglang{R}~code for replication of our results is
available at \url{http://semtools.R-Forge.R-project.org/}.

\bibliography{refs}
\end{document}

%% file: lmm_final-concordance.tex
\Sconcordance{concordance:lmm_final.tex:lmm_final.Rnw:%
1 80 1 1 0 4 1 1 18 208 1 1 25 103 1 1 10 5 1 1 13 1 2 52 1 1 18 1 2 %
344 1 1 28 7 1 1 18 1 2 8 1 1 18 1 2 10 1 1 18 1 2 9 1 1 18 1 2 8 1 1 %
18 1 2 8 1 1 18 1 2 8 1 1 18 1 2 5 1 1 57 33 1 1 2 1 0 4 1 3 0 1 2 21 1 %
1 2 1 0 2 1 3 0 1 2 22 1 1 3 2 0 1 1 1 3 1 0 1 3 7 0 1 2 6 1 1 2 1 0 1 %
1 1 3 1 0 1 2 1 0 1 2 4 0 1 3 7 0 1 2 13 1 1 3 2 0 4 1 1 3 1 0 1 2 1 0 %
1 2 1 0 1 1 3 0 1 3 4 0 1 2 11 1 1 3 5 0 1 3 7 0 1 2 44 1 1 2 1 0 2 1 3 %
0 1 2 13 1 1 9 21 1 1 53 129 1}